# Simple Flow Rules for Three-Phase Viscoplastic Materials

Frank Montheillet[1,a *] and David Piot[2,b]

[1,2]Mines Saint-Etienne, Univ Lyon, CNRS, UMR 5307 LGF

158 cours Fauriel, 42023 Saint-Etienne, France

[a]montheil@emse.fr, [b]david.piot@mines-stetienne.fr

**Abstract.** Noting that there is very little literature on the topic, a first analytical approach is proposed in this work for estimating the viscosity-like parameter of three-phase viscoplastic materials. In a first part, the conditions of application and the consequences of the three classical averaging equations involving the strain rates, the stresses and the power are reviewed for 2-phase mixtures and extended to three phases. The classical static and Taylor bounds as well as the heuristic "Iso-strain rate" assumption are analyzed. An extension of the Mori-Tanaka estimation to the three-phase case is then proposed for viscoplastic linear constituents. If the volume fraction of one of the phases (inclusions) is very low, in particular when its viscosity tends towards zero or infinity, fully analytical results are presented, which provides an extension of the classical dilute model.

**Introduction**

Stress-strain rate dependence of two-phase viscoplastic metal alloys or composites has been widely investigated in the past. Lower and upper bounds are available, as well as simple estimations for the viscosity-like (consistency) parameter of the mixture. In some situations, however, *three* phases co-exist in the material, for example in metal alloys where a two-phase matrix contains a dispersion of hard (*e.g.*, oxides) or soft (*e.g.*, lead) particles. Mixtures of three phases in roughly equal volume fractions can also be conceived. Very little literature is presently available on the topic, even for elastic components [1]. In this paper, a first analytical approach is proposed based on mere extensions of the classical Taylor, static, Iso-work, and Mori-Tanaka assumptions. Purely viscoplastic materials without strain-hardening are considered, *i.e.*, with flow rule $\sigma = k(\dot{\varepsilon}/\dot{\varepsilon}_0)^m$, where *k* is a viscosity-like parameter (with dimension of a stress), *m* the strain rate sensitivity, and $\dot{\varepsilon}_0 = 1\,s^{-1}$. The uniaxial *simple rules of mixture* considered in the following basically assume that strain rates and stresses are uniform over each phase domain, consisting of a single grain or a cluster of grains. Furthermore, they are identical in all domains of a given phase. For a multiphase composite, the calculations are based on the following three averaging equations:

$$\begin{cases} (a) & \sum_i f_i \dot{\varepsilon}_i = \dot{\varepsilon} \\ (b) & \sum_i f_i \sigma_i = \sigma \\ (c) & \sum_i f_i \sigma_i \dot{\varepsilon}_i = \sigma \dot{\varepsilon} \end{cases} \qquad (1)$$

which state that the macroscopic strain rate, flow stress, and power are the arithmetic averages of their microscopic counterparts [2]. Equation (1c) is commonly referred to as Hill's lemma. If the exponent *m* is identical for all phases, the three equations are homogeneous with respect to the strain rates, which means that the mixture follows a power law with the same exponent. The prescribed overall strain rate $\dot{\varepsilon}$ can thus be set to $1\,s^{-1}$ and the above system reduces to:

$$\begin{cases}(a) & \sum_i f_i \dot{\varepsilon}_i = 1 \\ (b) & \sum_i f_i k_i \dot{\varepsilon}_i^m = k \\ (c) & \sum_i f_i k_i \dot{\varepsilon}_i^{m+1} = k\end{cases} \qquad (2)$$

where $k$ is the viscosity-like parameter of the mixture. In the above equations, the volume fractions $f_i$ are of course related to each other by the relationship $\sum_i f_i = 1$.

For a mixture of $n$ phases, $n+1$ unknowns (the $\dot{\varepsilon}_i$ and $k$) are related by the three equations. In the case of a two-phase material, one thus gets three equations for three unknowns and the system can be solved by multiplying (a) by (b) and equaling to (c), which leads to:

$$(k_1 \dot{\varepsilon}_1^m - k_2 \dot{\varepsilon}_2^m)(\dot{\varepsilon}_1 - \dot{\varepsilon}_2) = 0 \qquad (3)$$

This equation highlights the two classical solutions referred to as Taylor or uniform strain rate model ($\dot{\varepsilon}_1 = \dot{\varepsilon}_2 = 1$) and static or uniform stress model ($k_1 \dot{\varepsilon}_1^m = k_2 \dot{\varepsilon}_2^m = k$), which are known to provide upper and lower bounds for the overall $k$, respectively.

**First Simple Flow Rules for a Three-Phase Mixture**

In the presence of three phases, only three equations are available for four unknowns ($\dot{\varepsilon}_1, \dot{\varepsilon}_2, \dot{\varepsilon}_3, k$). The system (2) is therefore indeterminate. It is nevertheless easy to check that the classical Taylor solution is still valid, with:

$$\begin{aligned} k_T &= f_1 k_1 + f_2 k_2 + f_3 k_3 \\ \dot{\varepsilon}_1 &= \dot{\varepsilon}_2 = \dot{\varepsilon}_3 = 1 \end{aligned} \qquad (4)$$

which is independent of the strain rate sensitivity $m$. In the same way, the static assumption verifies the equations (2) and gives:

$$\begin{aligned} \frac{1}{k_S^{1/m}} &= \frac{f_1}{k_1^{1/m}} + \frac{f_2}{k_2^{1/m}} + \frac{f_3}{k_3^{1/m}} \\ \dot{\varepsilon}_i &= \frac{\frac{1}{k_i^{1/m}}}{\sum_{j=1,3} \frac{f_j}{k_j^{1/m}}} = \left(\frac{k_S}{k_i}\right)^{1/m} \end{aligned} \qquad (5)$$

A purely heuristic alternative has long been proposed by Bouaziz and Buessler for elastic materials [3] and adapted to viscoplasticity by Montheillet and Damamme [4]. The method is based on the assumption of equal power of deformation in the phases of the mixture. In the case of three phases, this amounts to:

$$k_1 \dot{\varepsilon}_1^{m+1} = k_2 \dot{\varepsilon}_2^{m+1} = k_3 \dot{\varepsilon}_3^{m+1} \qquad (6)$$

which, when associated with the initial system (2) gives five equations for only four unknowns. In the two-phase case, the equivalent assumption also leads to one equation more than the number of unknowns. Equation (2c) is then generally discarded (even though it expresses the equality of internal and external powers). Doing the same in the present case, a system of four equations for four unknowns is obtained:

$$\begin{cases} (a) \quad f_1\dot{\varepsilon}_1 + f_2\dot{\varepsilon}_2 + f_3\dot{\varepsilon}_3 = 1 \\ (b) \quad f_1 k_1 \dot{\varepsilon}_1^m + f_2 k_2 \dot{\varepsilon}_2^m + f_3 k_3 \dot{\varepsilon}_3^m = k \\ (c) \text{ and } (d) \quad k_1 \dot{\varepsilon}_1^{m+1} = k_2 \dot{\varepsilon}_2^{m+1} = k_3 \dot{\varepsilon}_3^{m+1} \end{cases} \tag{7}$$

which can be solved analytically to give:

$$k_{Iso-\dot{w}} = \frac{\sum_{j=1,3} f_j k_j^{1/(m+1)}}{\left[\sum_{j=1,3} \frac{f_j}{k_j^{1/(m+1)}}\right]^m} \tag{8}$$

and:

$$\dot{\varepsilon}_i = \frac{\frac{1}{k_i^{1/(m+1)}}}{\sum_{j=1,3} \frac{f_j}{k_j^{1/(m+1)}}} \quad (i = 1, 2, 3) \tag{9}$$

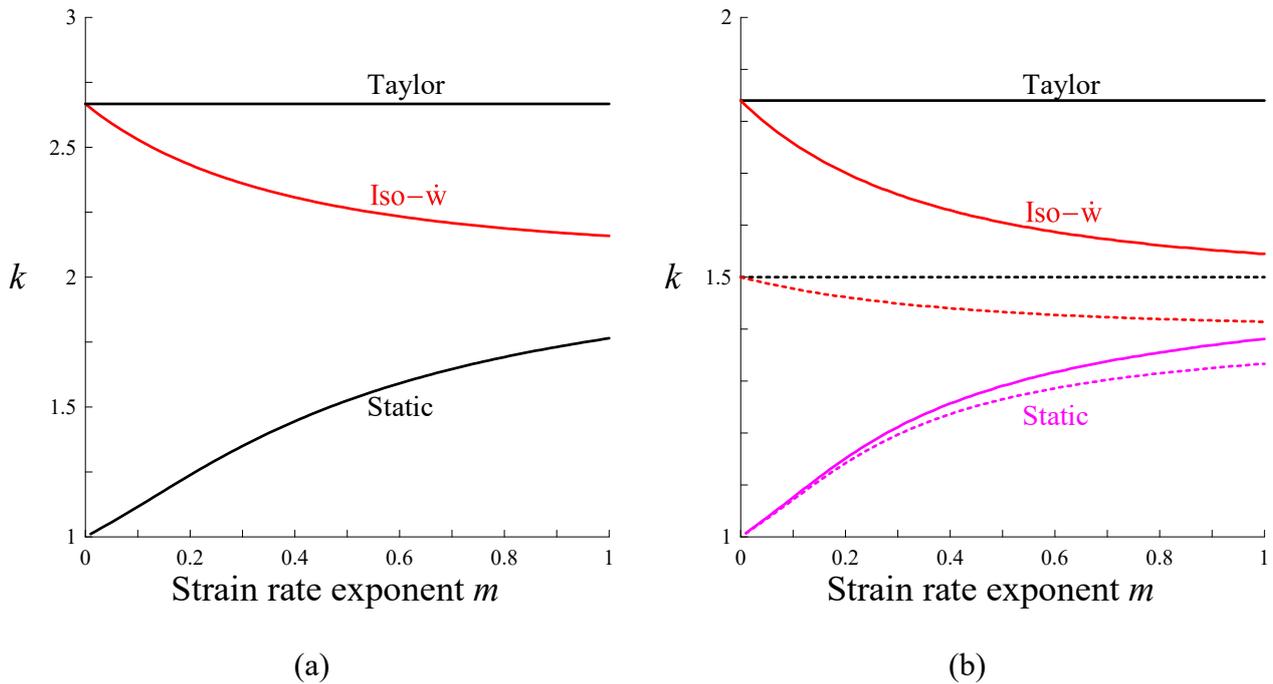

**Figure 1.** Predicted values of the overall viscosity-like parameter by the Taylor, static, and Iso-$\dot{w}$ assumptions as a function of the strain rate exponent $m$: (a) equivolumic mixture of three phases with $k_1 = 5$, $k_2 = 1$, $k_3 = 2$; (b) mixture of three phases with $k_1 = 10$, $f_1 = 0.04$ (inclusions), $k_2 = 1$, $f_2 = 0.48$, and $k_3 = 2$, $f_3 = 0.48$. Comparison is made with an equivolumic mixture of the same phases 1 and 2 without inclusions (broken lines)

The above results are illustrated in Figs 1a and b. Fig. 1a pertains to the case of an equivolumic ($f_1 = f_2 = f_3 = 1/3$) mixture of three phases. It shows that the Taylor and static bounds are very far from each other, especially for low $m$ values. The Iso-$\dot{w}$ prediction falls between the two bounds, which has been theoretically derived in [4]. In Fig. 1b, phase 1 with low volume fraction consists of inclusions embedded in a matrix made of a mixture of phases 2 and 3. The case of the same matrix without inclusions in shown in broken lines, which highlights the hardening effect of the inclusions. When $k_1 \to +\infty$ (non-deformable inclusions), while the Taylor and Iso-$\dot{w}$ predicted $k$ values tend to infinity, $k_S \to k_2 k_3 /(f_2 k_3^{1/m} + f_3 k_2^{1/m})^m$, which remains finite.

The Iso-$\dot{w}$ *local* strain rates are given by Eq. (9), from which the *local* flow stresses are derived, according to the power law $\sigma_i = k_i \dot{\varepsilon}_i^{\,m}$ for the three phases $i = 1, 2, 3$. Example of results are shown in Figs 2a and b for the same equivolumic mixture as in Fig. 1a.

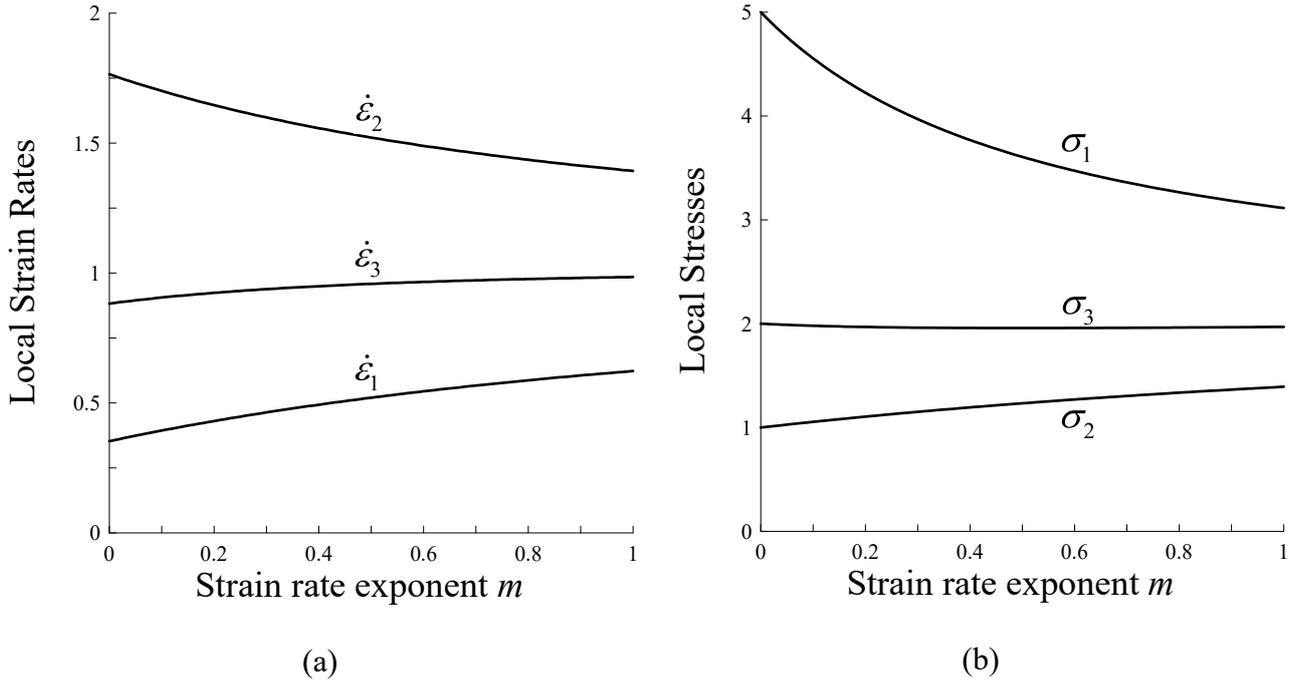

**Figure 2.** Iso-$\dot{w}$ predictions of the local strain rates (a) and flow stresses (b) in the three phases of the equivolumic mixture of Fig. 1a ($k_1 = 5$, $k_2 = 1$, $k_3 = 2$)

**Extension of the Mori-Tanaka Model to Three Phases**

In cases similar to that shown in Fig. 1b, where one of the phases clearly consists of inclusions in a mixture of the two others, an extension of the Mori-Tanaka approach is more relevant than the above models. In addition to the two averaging equations (2a) and (2b), a *localization equation* is introduced, which gives the strain rate $\dot{\varepsilon}_I$ of the inclusions as a function of the strain rate $\dot{\varepsilon}_M$ of the matrix. For a spherical linearly viscoplastic inclusion, it takes the simple form [5]:

$$\dot{\varepsilon}_I = \frac{5}{2\Sigma + 3} \dot{\varepsilon}_M \tag{10}$$

where $\Sigma = k_I / k_M$ is the ratio of the two viscosities. Numerical relations and analytical approximations similar to (10) are available for non-linear and/or non-spherical inclusions, which will not be considered here. Combining Eqs (2a), (2b), and (10) leads to:

$$k = \frac{(3 f_I + 2) k_I + 3(1 - f_I) k_M}{2(1 - f_I) k_I + (2 f_I + 3) k_M} k_M \tag{11a}$$

$$\dot{\varepsilon}_I = \frac{5k_M}{2(1-f_I)k_I + (2f_I + 3)k_M} \tag{11b}$$

$$\dot{\varepsilon}_M = \frac{2k_I + 3k_M}{2(1-f_I)k_I + (2f_I + 3)k_M} \tag{11c}$$

where $f_I$ is the volume fraction of inclusions.

For a three-phase material, calling 1 the inclusion phase with volume fraction $f_1$ and viscosity $k_I = k_1$, the matrix is a mixture of phases 2 and 3 with volume fraction $f_2 + f_3 = 1 - f_1$ (Fig. 3).

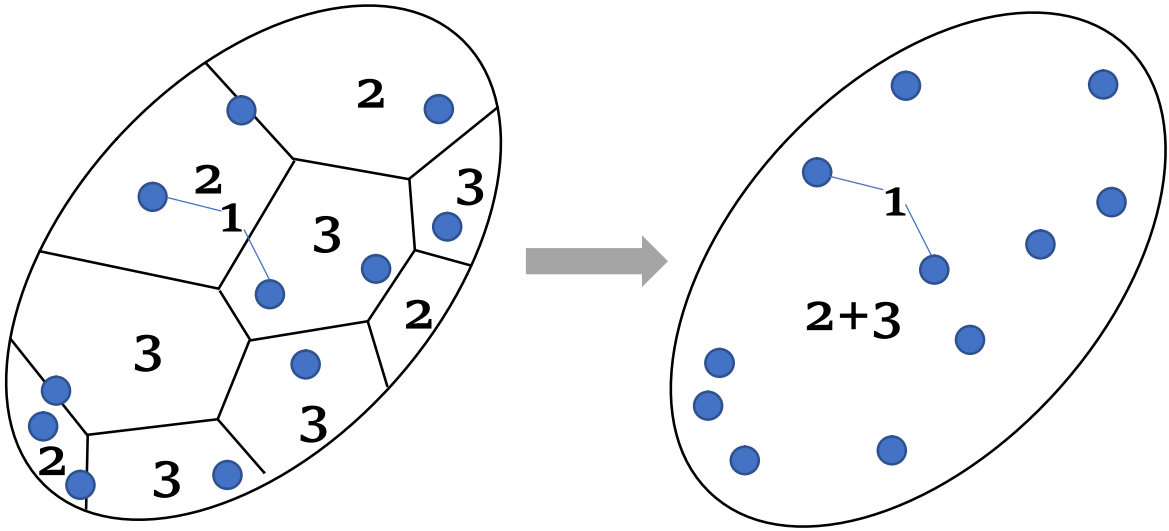

**Figure 3.** The three-phase material where phase 1 consists of inclusions in a mixture of phases 2 and 3. It is replaced by a two-phase material where the inclusions are embedded in a uniform matrix 2+3)

A simple way for estimating its viscosity $k_M = k_{23}$ is to use the Taylor assumption, which gives $k_{23} = (f_2 k_2 + f_3 k_3)/(1 - f_1)$. First results of this approach are displayed in Fig. 4, which shows the influence of the inclusion volume fraction $f_1$ on the overall viscosity $k$, for both "hard" and "soft" inclusions. For non-deformable inclusions ($k_1 \to \infty$), e.g. oxides, the above equations give the limit behaviour:

$$k \to \frac{3f_1 + 2}{2(1 - f_1)} \frac{f_2 k_2 + f_3 k_3}{1 - f_1} \tag{12}$$

which, for small volume fractions of inclusions ($f_1 \to 0$) becomes $k \to (1 + 5f_1/2)k_{23}$ or else $k \to (1 + 7f_1/2)(f_2 k_2 + f_3 k_3)$. In the first form, we recognize the classical equation of the dilute model, first obtained by Einstein in 1911. In the same way, for zero-viscosity (but still incompressible) inclusions ($k_1 = 0$), e.g. liquid lead, the overall viscosity takes the value:

$$k = \frac{3(1 - f_1)}{2f_1 + 3} \frac{f_2 k_2 + f_3 k_3}{1 - f_1} \tag{13}$$

which, for small volume fractions of inclusions ($f_1 \to 0$) becomes $k = (1 - 5f_1/3)k_{23}$ or else $k = (1 - 2f_1/3)(f_2 k_2 + f_3 k_3)$. The first formulation again coincides with the dilute model. These two limiting cases are also reported in Fig. 4.

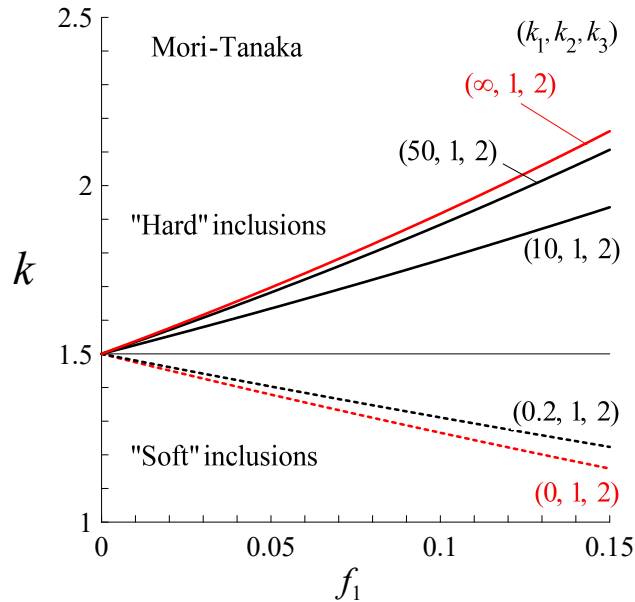

**Figure 4.** Influence of the volume fraction of inclusions (viscosity $k_1$) on the overall viscosity of a three-phase mixture. Solid and broken lines correspond to "hard" and "soft" inclusions, respectively. The values of the three viscosities ($k_1, k_2, k_3$) are given in brackets, including the two limiting cases of non-deformable and zero-viscosity inclusions

## Concluding Remarks

It would be unrealistic to expect to find enough experimental data (at least the viscosities and volume fractions of the three phases) to validate the above tentative theoretical predictions in detail. We can nevertheless hope that the overall trends and order of magnitudes can be verified in the near future.

In addition, obvious improvements or new alternatives will be introduced. So, in the case of roughly equivolumic three-phase materials, a self-consistent approach is likely to provide more accurate predictions. In the presence of inclusions embedded in a two-phase matrix, the consistency of the latter could be estimated by methods other than the mere Taylor bound.